\documentclass[12pt]{article}
\begin{document}
\title
{Bosonic p-brane and A--D--M decomposition}
\author
{ Rabin Banerjee\footnote{rabin@bose.res.in}\\
S. N. Bose National Centre for Basic Sciences \\
JD Block, Sector III, Salt Lake City, Kolkata -700 098, India
\and Pradip Mukherjee\footnote{pradip@bose.res.in} {}\footnote{Also Visiting Associate, S. N. Bose National Centre for Basic Sciences, JD Block, Sector III, Salt Lake City, Calcutta -700 098, India}
\\and\\
 Anirban Saha\\
Department of Physics, Presidency College\\
86/1 College Street, Kolkata - 700 073, India}
\maketitle
\abstract
{ A master action for the bosonic p-brane, interpolating between the Nambu--Goto and Polyakov formalisms, is discussed. The fundamental arbitrariness of extended structures (p-brane) embeded in space time manifold has been exploited to build an independent metric in the brane world volume. The cosmological term for the generic case follows naturally in the scheme. The dynamics of the structure leads to a natural emergence of the A--D--M like split of this world volume. The role of the gauge symmetries vis-\`{a}-vis reparametrization symmetries is analyzed by a constrained Hamiltonian approach. }

\noindent {\bf PAC codes:} 11.15.-q, 11.25.-w 

\section{Introduction}
String theory is considered as the candidate theory to unite all fundamental interactions including gravity. For uniqueness and consistency of different perturbative string theories, extended structures like membranes \cite{ct} are required to be introduced with an independent physical status \cite{mr}. 
A generic element of this class is a p-dimensional spatially extended structure called the p-brane. Apart from the symmetries of the space-time in which the brane is embeded it also carries various symmetries associated with reparametrisation of the world volume. These symmetries have been discussed extensively in the literature. However, all these studies are based on either the Nambu--Goto (N--G) \cite{ct, fl, za, skb, fitz} or the Polyakov-type \cite{fuji} actions and a systematic analysis of the mutual correspondence between the different formalisms is lacking. Aspects of symmetries continue to be of fundamental importance in the study of dynamics of the brane and it is indeed crucial to understand these issues from a unified point of view. The present paper will be devoted to the analysis of the basic world volume symmetries of a bosonic p-brane from such a unified approach. Implications and consequences of the symmetries will also be discussed.

  The p-brane sweeps out a $p+1$ dimensional world volume in the embedding space time. The dynamics of the brane can be analysed from different action formalisms. In the N--G description the physical action is prescribed solely in terms of the space time coordinates of the brane, taken as independent fields. Alternatively, in the Polyakov action formalism the metric in the world volume is considered as a collection of independent fields in addition to the usual space time coordinates. The equivalence of these two approaches is usually established by starting from the Polyakov action and solving out the independent metric in favour of the space time coordinates. In the present paper, on the contrary, we address the reverse problem by demonstrating how the independent metric can be generated by exploiting the reparametrization symmetry of the NG action for the p-brane. An intermediate step is the construction of an interpolating action. Such actions were earlier introduced
for discussing various aspects of symmetries and noncommutativity in the case of strings and membranes \cite{rbbcsg, bms, rbk}.
However the methods used were specific to the particular choice of $p=1 (strings), p=2 (membranes)$ only, which do not admit a generalisation to the arbitrary p-case that is essential for the present analysis. 
The interpolating action is based on the first class constraints of  
the NG theory. The independent metric will be generated from the corresponding Lagrange multipliers enforcing these constraints. This reveals a deep connection of the metric components with the gauge symmetries of the brane.
The mismatch between the number of independent gauge degrees of freedom and the number of independent metric elements brings out the arbitrariness, inherent in the Polyakov formulation, explicit in our construction. Fixing the arbitrariness in terms of the embedding makes the transition to the Polyakov form complete. Notably, the cosmological term emerges as a logical consequence of our analysis. 

  The process of introducing the independent metric in the world volume through the interpolating action formalism has a very interesting outcome. First class constraints of the N--G theory generate temporal development and also shifts in the space like directions. The independent metric constructed with the help of the Lagrange multipliers enforcing these constraints naturally emerge with a decomposition of the $(p+1)$-dimensional metric into the $(p)$-dimensional spatial part plus the multipliers which are the analogues of the lapse and shift variables of general relativity. Indeed, the metric generated in our formalism appears in a canonical form which is shown to be identical with the famous Arnowitt--Deser--Misner (A--D--M)
representation in general relativity. In other words our analysis provides a genesis of the A-D-M representation from a string theoretic perspective. 


The interpolating action formalism is based on the gauge symmetries of the theory. These gauge degrees of freedom actually correspond to the invariance under reparametrisation of the brane world volume. Considering the pivotal role played by the gauge symmetries in our analysis, it is only natural to undertake a thorough investigation of these symmetries vis-\`{a}-vis the reparametrisation symmetries. In this paper we address this problem using a Hamiltonian method of abstracting the independent gauge parameters introduced earlier in the literature \cite{brr}. 

 The organisation of the paper is as follows. In section 2 we introduce the interpolating action for the generic p-brane and discuss its passage to the N--G and the Polyakov forms. Specifically, the method of obtaining the independent metric components from the fields of the interpolating theory has been elaborated. Various consistency conditions have been deduced which are used in the sequel. In section 3 the emergence of the A--D--M decomposition of the p-brane world volume from our construction has been indicated. Section 4 discusses a comprehensive analysis of the gauge symmetries of the interpolating action and its parallel with the reparametrisation invariance of the theory. We conclude in section 5.

\section{Interpolating action for the bosonic p-brane}
The p-brane is a $p$ dimentional object which sweeps out a $\left(p+1\right)$ dimensional world volume parametrised by $\tau$ and $\sigma_{a}$. The index $a$ run from $1$ to $p$. Henceforth these parameters are collectively referred as $\xi_{i}\left(\xi_{0}= \tau, \xi_{a}= \sigma_{a}\right)$. The N--G action of bosonic p-brane is the integrated proper area of this world volume: 
\begin{eqnarray}
S_{\mathrm{NG}} = - \int d^{p+1}\xi \sqrt{-h}
\label{ngaction}
\end{eqnarray}
where $h$ is the determinant of the induced metric
\begin{equation}
h_{ij} = \partial_{i} X^{\mu}\partial_{j}X_{\mu}
\label{indhij}
\end{equation}
Note that we have kept the p-brane tension implicit. 
The canonical momenta conjugate to $X_{\mu}$ are 
\begin{eqnarray}
\Pi_{\mu} = \frac{\bar{h}}{\sqrt{-h}}\left\{\partial_{0}X_{\mu} - \partial_{a}X_{\mu} {\bar{h}}^{ab} h_{0b}\right\}
\label{momentumX}
\end{eqnarray}
where ${\bar{h}} $ is the determinant of the matrix $h_{ab}$.  Also ${\bar{h}}^{ab}$
{\footnote
{Note that $\bar h^{ab}$ is different from the space part of $h^{ij}.$}} 
is the inverse of $h_{ab}$.
The primary constraints following from (\ref{momentumX})
are,  
\begin{eqnarray}
\Omega_{0} = \frac{1}{2}\left(\Pi^2 + \bar{h}\right) \approx 0; \quad
\Omega_{a} = \Pi_{\mu}\partial_{a}X^{\mu} \approx 0 
\label{constraints}
\end{eqnarray}
The nontrivial Poission's bracket of the theory are given by 
\begin{eqnarray}
 \{X^{\mu}\left(\tau,\xi\right),
 \Pi_{\nu}\left(\tau,\xi^{\prime}\right)\} = \eta_{\nu}^{\mu}
 \delta\left(\xi - \xi^{\prime}\right)
\label{XPicommutator}
\end{eqnarray}
Using these Poisson brackets it is easy to work out the algebra of the
constraints
\begin{eqnarray}
\left\{ \Omega_{0}\left( \xi \right), \Omega_{0}\left( \xi^{\prime} \right)
\right\} &=& 4 \left[ \bar {h}\left( \xi \right)\bar {h}^{ab}
\left( \xi \right)\Omega_{b}\left( \xi \right) \right.\nonumber \\
&& + \left.\bar {h}\left( \xi^{\prime} \right)\bar {h}^{ab}\left( \xi^{\prime} \right)
\Omega_{b}\left( \xi^{\prime} \right) \right]
\partial_{a}\left(\delta\left(\xi - \xi^{\prime}\right)\right)\nonumber \\
\left\{ \Omega_{a}\left( \xi \right), \Omega_{0}\left( \xi^{\prime} \right)
\right\} &=& \left[\Omega_{0}\left( \xi \right) + \Omega_{0}\left( \xi^{\prime}
\right)\right] \partial_{a}\left(\delta\left(\xi - \xi^{\prime}\right)\right)
\nonumber\\
\left\{ \Omega_{a}\left( \xi \right), \Omega_{b}\left( \xi^{\prime} \right)
\right\} &=& \left[\Omega_{b}\left( \xi \right) \partial_{a}\left(\delta
\left(\xi - \xi^{\prime} \right)\right)
- \Omega_{a}\left( \xi^{\prime} \right)  \partial_{b}^{\prime} \left(\delta
\left(\xi - \xi^{\prime} \right)\right)\right] 
\label{constraintalgebra}
\end{eqnarray}

   Clearly, the Poisson brackets between the constraints (\ref{constraintalgebra}) are weakly involutive so that the set (\ref{constraints}) is first class. 
Since the p-brane action (\ref{ngaction}) possesses reparametrization invariance, the canonical Hamiltonian following from the action vanishes. Thus the total Hamiltonian is only a linear combination of the constraints (\ref{constraints}):
\begin{equation}
{\cal{H}}_{T} = - \frac{\lambda_{0}}{2}\left(\Pi^2 + \bar{h}\right)  - \lambda^{a}\Pi_{\mu}\partial_{a}X^{\mu} 
\label{totalHamiltnian}
\end{equation}
In the above expression $\lambda_{0}$ and $\lambda^{a}$ are the Lagrange multipliers.

The Polyakov action for the p-brane is given by,
\begin{equation}
S_{\mathrm{P}} = -\frac{1}{2}\int d^{p+1}\xi{\sqrt - g}\left\{g^{ij} \partial_{i} X^{\mu}\partial_{j} X_{\mu} -\left(p - 1\right)\right\}
\label{polyakovaction}
\end{equation}
The metric $g_{ij}$ are now considered as independent fields. The equations of motion for $g_{ij}$ are 
\begin{equation}
g_{ij} = h_{ij}
\label{polyeqnmotion}
\end{equation}
Substituting these in (\ref{polyakovaction}) we retrieve the N--G form (\ref{ngaction}). Note the cosmological term ${\sqrt - g}\left(p - 1\right)$ in the action. For $p = 1$ this term vanishes. We thus observe that the presence of the cosmological term is characteristic of the higher branes as opposed to the strings. The reason for this difference is the Weyl invariance of the string which is not shared by the higher branes. In our action level construction this cosmological term will emerge systematically. 

 We now come to the construction of an interpolating action for the p-brane. The first step  is to consider the Lagrange multipliers as independent fields and write an alternative first order Lagrangian for the p-brane
\begin{equation}
{\cal{L}}_I = \Pi_{\mu}\dot{X}^{\mu} - \cal{H}_{T}
\label{firstorderLagrangean}
\end{equation}
The equation of motion for $\Pi_{\mu}$ following from the Lagrangian
(\ref{firstorderLagrangean}) is
\begin{equation}
\Pi_{\mu} = -\frac{\dot{X_{\mu}}+\lambda^a\partial_{a}X_{\mu}}{\lambda_{0}}
\label{Pieqnmotion}
\end{equation}
Substituting $\Pi_{\mu}$ in (\ref{firstorderLagrangean}) we get the interpolating Lagrangian
\begin{eqnarray}
{\cal{L}}_I = -\frac{1}{2\lambda_{0}}\left[\dot{X^{\mu}}\dot{X_{\mu}}
                 + 2\lambda^{a}\dot{X}_{\mu}\partial_{a}X^{\mu}
                 +\lambda^{a}\lambda^{b}\partial_{a}X^{\mu}\partial_{b}X_{\mu}\right]
                 +\frac{\lambda_{0}}{2}\bar{h}
\label{interpolatingLagrangean}
\end{eqnarray}
for the p-brane. 

 The Lagrangian (\ref{interpolatingLagrangean}) has been referred to as the interpolating Lagrangian because this can be reduced to either the N--G or the Polyakov form of the p-brane action. Let us first discuss the passage to the N--G form. From the interpolating Lagrangian it is easy to write down the equations of motion for $\lambda_{0}$ and $\lambda^{a}$ :
\begin{eqnarray}
\lambda_{0}{}^{2} &=& \frac{-h}{{\bar{h}}^{2}} \nonumber \\
\lambda^{a} &=& -h_{0b}\bar h^{ba}
\label{LambdaRhoeqnmotion}
\end{eqnarray}
From the first equation of (\ref{LambdaRhoeqnmotion}) $\lambda_{0}$ is determined modulo a sign. This can be fixed by demanding the consistency of (\ref{momentumX}) with (\ref{Pieqnmotion}), the equation of motion for $\Pi_{\mu}$ following from the first order Lagrangian (\ref{firstorderLagrangean}). Thus we have 
\begin{equation}
\lambda_{0} = -\frac{\sqrt{-h}}{\bar{h}}
\label{Lambda}
\end{equation}
Substituting $\lambda^a$ and $\lambda_{0}$ in (\ref{interpolatingLagrangean}) we retrieve the Nambu--Goto action (\ref{ngaction}). 

The reduction of the interpolating Lagrangian to the Polyakov form of the p-brane action is non-trivial. 
In deriving the interpolating Lagrangean from the N--G theory we have promoted the $\left(p+1\right)$ Lagrange 
multipliers as independent fields. Note that in the Polyakov action the extra degrees of freedom is more than this number. The precise size of the mismatch is $\left(p\right)\left(p+1\right)/2$. We thus observe that the interpolating action is a less redundant description than the Polyakov action. So to make the transition from the interpolating Lagrangean to the Polyakov form we require to introduce just as many independent fields. This can be done by including an arbitrary spatial part $G^{ab}$ in ${\cal{L}}_I$, which has the right number of independent components. We therefore modify the interpolating Lagrangian (\ref{interpolatingLagrangean}) for the p-brane in the following way 
\begin{eqnarray}
{\cal{L}}_I & = & -\frac{1}{2\lambda_{0}}\left[\dot{X^{\mu}}\dot{X_{\mu}}
                 + 2\lambda^a\dot{X_{\mu}}\partial_a X^{\mu} + \left(\lambda^a\lambda^b \partial_a X^{\mu}\partial_b X_{\mu} 
                  -   \lambda_{0}{}^{2}\bar{G} G^{ab}  \partial_{a}X_{\mu} \partial_{b} X^{\mu}\right)\right]\nonumber \\
                  &&- \frac{\lambda_{0}}{2}\left(\bar{G} G^{ab}
                     \partial_{a} X_{\mu} \partial_{b} X^{\mu} - \bar{h}\right) 
\label{modintL}
\end{eqnarray}
where $\bar{G}$ is the determinant of $G_{ab}$ which is the inverse of the arbitrary matrix $G^{ab}$, $\left( a,b = 1, 2,... p\right)$. This specific choice of the arbitrary part will be convenient in the subsequent calculation. Observe that (\ref{modintL}) can be cast as 
\begin{equation}
{\cal{L}}_I = -\frac{1}{2}\sqrt{-g} g^{ij}\partial_i X_{\mu}
               \partial_j X^{\mu} - \frac{\lambda_{0}}{2}\left(\bar{G} G^{ab}
              \partial_{a} X_{\mu} \partial_{b} X^{\mu} - \bar{h}\right) 
\label{almostpoly}
\end{equation}
where 
\begin{equation}
g^{ij} = \left(-g\right)^{-\frac{1}{2}}
\left(\begin{array}{cc}
\frac{1}{\lambda_{0}} & \frac{\lambda^a}{\lambda_{0}} \\
\frac{\lambda^b}{\lambda_{0}} & \frac{\lambda^{a} \lambda^{b} - \lambda_{0}{}^2 \bar{G} G^{ab} }{\lambda_{0}} \\
\end{array}\right)
\label{idm}
\end{equation}
Here $g$ is the determinant of $g_{ij}$ which 
is the inverse of $g^{ij}$. This imposes stringent constraints on the 
construction (\ref{idm}). So its consistency 
must explicitly be examined. 
Observe that by exploiting the 
dynamics of the p-brane we are able to 
generate an independent metric on the world volume of the brane. 
The arbitrary function $G_{ab}$ signifies a fundamental 
elasticity in the spatial part of the metric. 
The Lagrangean (\ref{almostpoly}) is almost in the required Polyakov form except for the omission of the cosmological constant. Also there is an additional term which is not there in the Polyakov Lagrangean. It is precisely the consistency requirement of the construction (\ref{idm}) which identifies this extra piece in (\ref{almostpoly}) with the cosmological constant, provided we fix the elasticity in the embedding. The validity of these assertions will be demonstrated now.
 
 From the identification (\ref{idm}) we get after a straightforward calculation that
\begin{equation}
{\rm{det}} g^{ij} = \left(-1\right)^{p}\frac{\lambda_{0}{}^{\left(p-1\right)}}{\left(\sqrt{-g}\right)^{\left(p+1\right)}}{\rm{det}}\left(\bar G G^{ab}\right)
\label{detidm}
\end{equation}
But we require ${\mathrm{det}} g^{ij}$ = $ g^{-1} $. Comparing, we get the
condition
\begin{equation}
\lambda_{0}{}^{\left(p-1\right)} = \left(-1\right)^{\left(1-p\right)} \left(\frac{\sqrt{-g}}{{\bar G}}\right)^{\left(p-1\right)}
\label{consistancy1}
\end{equation}
Starting from our construction (\ref{idm}) one can solve for $\lambda_{0}$ and $\lambda^{a}$ as 
\begin{eqnarray}
\frac{1}{\lambda_{0}} = \sqrt{-g} g^{00}, \quad
\lambda^{a}= \frac{g^{0a}}{g^{00}}
\label{c1}
\end{eqnarray}
Using (\ref{c1}) in (\ref{idm}), we get after a few steps 
\begin{equation}
{G}^{ab} = \frac{g}{\bar G}\left(g^{ab} g^{00} - g^{0a} g^{0b}\right)
\label{spatialidm}
\end{equation}
Inverting ${G}^{ab}$ we arrive at
\begin{equation}
 G_{ab} = \left(\frac{g g^{00}}{{\bar G}}\right)g_{ab}
\label{inversespatialidm}
\end{equation}
From (\ref{spatialidm}) we obtain after some calculations  
\begin{equation}
{\mathrm {det}}{G}^{ab} = \left(\frac{g}{\bar G}\right)^{p}{\mathrm {det}}{g}^{ij} \left(g^{00}\right)^{p-1}  
\label{detinversespatialidm}
\end{equation}
But, by definition, ${\mathrm {det}}{G}^{ab} = 1/\bar{G}$. Using this in (\ref{detinversespatialidm}) we find,
\begin{equation}
{\bar G}^{\left(p-1\right)} = \left(g g^{00}\right)^{\left(p-1\right)}
\label{consistancy2}
\end{equation}
There is an apparent ambiguity of sign in determining $\bar {G}$ from (\ref{consistancy2}) when $p$ is odd. For now we take the positive solution for all $p$. Then from (\ref{inversespatialidm})
\begin{equation}
G_{ab} = g_{ab}
\label{elasticity}
\end{equation}
The consistency requirment thus restricts the arbitrariness of $G_{ab}$ through (\ref{elasticity}).
We use (\ref{consistancy1}) and (\ref{elasticity}) to express the $\sqrt{-g}$ factor in terms of the $\bar{G}$ and $\lambda_{0}$ as {\footnote{
Note that for odd $p$ another sign ambiguity appears here. This is actually related with the corresponding uncertainty about sign stated above. We shall explore the connection subsequently.
}}

\begin{equation}
\sqrt{-g} = -\lambda_{0} \bar{G} = - \lambda_{0} {\rm det} g_{ab}
\label{theg}
\end{equation}
Finally, we discuss the reduction of the interpolating action to the Polyakov form. Note that the spatial part of the metric $g^{ij}$ is still remaining arbitrary. Also no attention has so far been paid to the background space time in which the brane is embedded. We now propose the rigid structure 
\vskip -0.50cm
\begin{equation}
g_{ab} = h_{ab}
\label{rigidity}
\end{equation}
\vskip -0.25cm
In this connection it may be observed that this is just the spatial part of (\ref{polyeqnmotion}) which is required to demonstrate the equivalence of the Polyakov form with the N--G. Now equation (\ref{rigidity}), along with (\ref{elasticity}), imposes 
\begin{equation}
G_{ab} = h_{ab}
\label{result}
\end{equation}
Plugging it in the Lagrangean (\ref{almostpoly}) and using (\ref{theg}) 
we find that the last term of (\ref{almostpoly}) is precisely 
equal to the cosmological constant occurring in the 
Polyakov action (\ref{polyakovaction}). This completes the reduction of the 
interpolating Lagrangian to the Polyakov form. The connection (\ref{rigidity}) 
fixes the brane in its embedding. 
A couple of interesting observations also follow from this. 
First, we can understand now the nature of 
the ambiguities of sign encountered above for odd $p$ more clearly.  
If we chose the opposite sign in (\ref{elasticity}) then we would have $G_{ab} = - h_{ab}$ and $\lambda_{0}$ should then be expressed from (\ref{consistancy1}) as $\lambda_{0} = \frac{\sqrt{-g}}{\bar{G}}$. Otherwise there would be contradiction with (\ref{Lambda}). Next, for $p = 1$ we find that the imposed rigidity admits a residual scale transformation. This is the well known Weyl invariance of the string.\\


\section{Emergence of A--D--M decomposition from the brane dynamics}
The interpolating action formalism enables us to introduce an independent metric in the world volume swept out by the N--G brane. The process depends crucially on the first class constraints of the theory. We have also clearly identified the arbitrariness in the spatial part of the metric. Our method thus introduces the metric in a very special way such that the world volume is decomposed into the $p$ dimensional spatial part along with the multipliers which generate temporal evolution  with shifts in the space-like direction. This decomposition of the metric is reminiscent of the A--D--M decomposition in geometrodynamics. Indeed, the connection with A--D--M decomposition was noticed earlier in the special cases of string and membrane \cite{bms}. We are now in a position to show how the A--D--M representation follows from the dynamics of the generic p-brane.
To see this we have to use (\ref{theg}) to first express the metric in terms of its arbitrary spatial part and the Lagrange multipliers only. The construction (\ref{idm}) then reduces to
\begin{equation}
g^{ij} = 
\left(\begin{array}{cc}
-\frac{1}{\lambda_{0}{}^{2}{\rm det} g_{ab}} & -\frac{\lambda^a}{\lambda_{0}{}^{2}{\rm det} g_{ab}} \\
-\frac{\lambda^b}{\lambda_{0}{}^{2}{\rm det} g_{ab}} & \left(\bar{g}^{ab} -\frac{\lambda^{a} \lambda^{b}}{\lambda_{0}{}^{2}{\rm det} g_{ab}}\right) \\
\end{array}\right)
\label{idmm}
\end{equation}
where $\bar{g}^{ab}$ is the inverse of the spatial metric $g_{ab}$\footnote{Note that as in the case of $\bar{h}^{ab}$, $\bar{g}^{ab}$ is also different from the spatial part of the identification matrix $g^{ij}$.}. 
In the A--D--M construction the metric $g^{ij}$ of physical space time is represented as 
\begin{equation}
g^{ij} = 
\left(\begin{array}{cc}
-\frac{1}{\left(N\right)^{2}} & \frac{N^a}{\left(N\right)^{2}} \\
\frac{\lambda^b}{\left(N\right)^{2}} & \left(\bar{g}^{ab} -\frac{N^{a} N^{b}}{\left(N\right)^{2}}\right) \\
\end{array}\right)
\label{idmm}
\end{equation}
where $N$ and $N^{a}$ are respectively the lapse and shift variables and $\bar{g}^{ab}$ is the inverse of the `metric' $g_{ab}$ on the spatial hypersurface. Using the correspondence 
\begin{eqnarray}
\left(N^{a}\right)\mapsto -\lambda^{a},\quad \mathrm {and}\quad N \mapsto \lambda_{0} \sqrt{{\rm det} g_{ab}}
\label{ADM5}
\end{eqnarray}
it is easy to convince oneself that the A--D--M decomposition of the brane volume emerges from our analysis. Note that in the correspondences (\ref{ADM5}) apart from the Lagrange multipliers only the  space part of the metric $g_{ij}$ is involved. The flexibility in $g_{ab}$ is apparent in our equation (\ref{elasticity}). Modulo this arbitrariness the lapse and shift variables are the fields $\lambda_{0}$ and $\lambda^{a}$ in our interpolating Lagrangean (\ref{almostpoly}). They in turn owe their existence to the constraints (\ref{constraints}) which are nothing but the superhamiltonian and supermomentum of the theory. Our interpolating Lagrangean (\ref{interpolatingLagrangean}) can thus be considered as the brane analog of the A--D--M formulation of geometrodynamics.

\section{Constraint structure and gauge symmetry}
The interpolating action formalism offers a composite scenario for discussing alternative actions of the p-brane. A thorough understanding of its gauge symmetries will thus be very much appropriate to our context. 
In this section we will discuss the gauge symmetries of the different actions and find their exact correspondence with the reparametrization invariances. Since our discussion will be centered on the interpolating action (\ref{interpolatingLagrangean}) let us begin with an analysis of its constraint structure. The independent fields in (\ref{interpolatingLagrangean}) are $X_{\mu}$, $\lambda_{0}$ and $\lambda^{a}$. Let the corresponding momenta be denoted by $\Pi_{\mu}$, $\Pi_{\lambda_{0}}$ and $\Pi_{\lambda^{a}}$ respectively. By definition
\begin{eqnarray}
\Pi_{\mu}& =&  -\frac{1}{\lambda_{0}}\left(\dot{X}_{\mu} +
               \lambda^{a} \partial_{a}X_{\mu}\right)\nonumber\\ 
\Pi_{\lambda_{0}}& = & 0 \nonumber\\
\Pi_{\lambda^{a}}& = & 0 
\label{intmomentum}
\end{eqnarray}
In addition to the Poisson brackets similar to (\ref{XPicommutator}) we now have
\begin{eqnarray}
\{\lambda_{0}\left(\tau,\xi\right),\Pi_{\lambda_{0}}\left(\tau,\xi^{\prime}\right)\} & = &
\delta\left(\xi - \xi^{\prime}\right)\nonumber\\
\{\lambda^{a}\left(\tau,\xi\right), \Pi_{\lambda^{b}}\left(\tau,\xi^{\prime}\right)\} & = & \delta\left(\xi - \xi^{\prime}\right) \delta^{a}_{b} \nonumber\\
\label{intcommutator}
\end{eqnarray}
The canonical Hamiltonian following from (\ref{interpolatingLagrangean}) is
\begin{equation}
{\cal{H}}_c = -\lambda^{a}\Pi_{\mu}\partial_{a}X^{\mu} 
              - \frac{\lambda_{0}}{2}\left(\Pi^2 +X^{\prime 2}\right) 
\label{intcanonicalH}
\end{equation}
which reproduces the total Hamiltonian of the NG action. From the definition of the canonical momenta we can easily identify the primary constraints
\begin{eqnarray}
\Pi_{\lambda_{0}}& \approx & 0\nonumber\\
\Pi_{\lambda^{a}}& \approx & 0 
\label{intprimaryconstraints}
\end{eqnarray}
Conserving these primary constraints we find that two new secondary constraints emerge. These are the constraints of equation (\ref{constraints}), as expected. The primary constraints of the NG action appear 
as secondary constraints in this formalism. No more secondary constraints are obtained. The system of constraints for the Interpolating Lagrangean thus comprises of the set (\ref{constraints}) and (\ref{intprimaryconstraints}). The Poisson brackets of the constraints of (\ref{intprimaryconstraints}) vanish within themselves. Also the PB of these with (\ref{constraints}) vanish. All the constraints are first class and therefore generate gauge transformations on ${\cal{L}}_I$ but the number of independent gauge parameters is equal to the number of independent primary first class constraints.  In the following analysis we will apply a systematic procedure \cite{brr}  of abstracting the most general local symmetry transformations of the Lagrangean. A brief review of the procedure will thus be appropriate.

    Consider a theory with first class constraints only. The set of constraints $\Omega_{a}$ is assumed to be classified as
\begin{equation}
\left[\Omega_{a}\right] = \left[\Omega_{a_1}
                ;\Omega_{a_2}\right]
\label{215}
\end{equation}
where $a_1$ belong to the set of primary and $a_2$ to the set of
secondary constraints. The total Hamiltonian is
\begin{equation}
H_{T} = H_{c} + \Sigma\lambda^{a_1}\Omega_{a_1}
\label{216}
\end{equation}
where $H_c$ is the canonical Hamiltonian and $\lambda^{a_1}$ are Lagrange multipliers enforcing the primary constraints. The most general expression for the generator of gauge transformations is obtained according to the Dirac conjecture as
\begin{equation}
G = \Sigma \epsilon^{a}\Omega_{a}
\label{217}
\end{equation}
where $\epsilon^{a}$ are the gauge parameters, only $a_1$ of which are independent. By demanding the commutation of an arbitrary gauge variation with the total time derivative,(i.e. $\frac{d}{dt}\left(\delta q \right) = \delta \left(\frac{d}{dt} q \right) $) we arrive at the following equations \cite{brr,htz}
\begin{equation}
\delta\lambda^{a_1} = \frac{d\epsilon^{a_1}}{dt}
                 -\epsilon^{a}\left(V_{a}^{a_1}
                 +\lambda^{b_1}C_{b_1a}^{a_1}\right)
                              \label{218}
\end{equation}
\begin{equation}
  0 = \frac{d\epsilon^{a_2}}{dt}
 -\epsilon^{a}\left(V_{a}^{a_2}
+\lambda^{b_1}C_{b_1a}^{a_2}\right)
\label{219}
\end{equation}
Here the coefficients $V_{a}^{a_{1}}$ and $C_{b_1a}^{a_1}$ are the structure
functions of the involutive algebra, defined as
\begin{eqnarray}
\{H_c,\Omega_{a}\} = V_{a}^b\Omega_{b}\nonumber\\
\{\Omega_{a},\Omega_{b}\} = C_{ab}^{c}\Omega_{c}
\label{2110}
\end{eqnarray}
Solving (\ref{219}) it is possible to choose $a_1$ independent
gauge parameters from the set $\epsilon^{a}$ and express $G$ of
(\ref{217}) entirely in terms of them. The other set (\ref{218})
gives the gauge variations of the Lagrange multipliers. It can be shown that
these equations are not independent conditions but appear as internal
consistency conditions. In fact the conditions (\ref{218}) follow from
(\ref{219}) \cite{brr}.

We begin the analysis with the interpolating action (\ref{interpolatingLagrangean}). Here the fields are $X^{\mu}, \lambda_{0}$ and $\lambda^{a}$. The set of constraints are given by (\ref{intprimaryconstraints}) along with (\ref{constraints}). All these constraints are first class. Denoting these by the set $\left\{\Psi_{k}\right\}$ we write the generator of the gauge transformations of (\ref{interpolatingLagrangean}) as 
\begin{equation}
G =\int d\xi \alpha_{k}\Psi_{k}
\label{gaugegenerator}
\end{equation}
where $\alpha_{k}$ are the gauge parameters.
We could proceed from this and construct the generator of gauge transformations from (\ref{gaugegenerator}) by including the whole set of first class constraints (\ref{constraints}, \ref{intprimaryconstraints}). Using (\ref{219}) the dependent gauge parameters can be eliminated. After finding the gauge generator in terms of the independent gauge parameters, the variations of the fields $X^{\mu}$, $\lambda_{0}$ and $\lambda^{a}$ can be worked out. However, looking at the intermediate first order form (\ref{firstorderLagrangean}) we understand that the variations of the fields $\lambda_{0}$ and $\lambda^{a}$ can be calculated alternatively, ( using (\ref{218})) from the NG theory where they appear as Lagrange multipliers. We adopt this alternative procedure. The generator of gauge transformations has already been given in (\ref{gaugegenerator}) where $\Omega_{i}$ now stands for the first-class constraints of the N--G theory, i.e. (\ref{constraints}).
The  variations of $ \lambda_{i}$ are obtained from (\ref{218})
\begin{equation}
\delta\lambda_{i}\left( \xi\right) = -  \dot \alpha_{i}
- \int d\xi^{\prime} d\xi^{\prime \prime} C_{kj}{}^{i}
\left(\xi^{\prime}, \xi^{\prime \prime}, \xi\right)
\lambda_{k}\left(\xi^{\prime}\right) \alpha_{j}
\left(\xi^{\prime \prime}\right)
\label{gaugevariation}
\end{equation}
where
$ C_{kj}{}^{i}\left(\xi^{\prime}, \xi^{\prime \prime}, \xi\right)$ are given by
\begin{equation}
\left\{ \Omega_{\alpha}\left(\xi\right),\Omega_{\beta}
\left(\xi^{\prime}\right)\right\} =  \int d\xi^{\prime \prime}
C_{\alpha \beta}{}^{\gamma}\left(\xi, \xi^{\prime}, \xi^{\prime
\prime}\right) \Omega_{\gamma}\left(\xi^{\prime \prime}\right)
\label{structurefunctions}
\end{equation}
Observe that the structure function $ V_{a}{}^{b}$ does not appear in (\ref{gaugevariation}) since $H_{c} = 0$ for the NG theory. The nontrivial structure functions $C_{\alpha \beta}{}^{\gamma}\left(\xi, \xi^{\prime}, \xi^{\prime \prime}\right)$ are obtained from the N--G constraint algebra (\ref{constraintalgebra}).
\begin{eqnarray}
C_{00}^{b} = 4\left[ \bar{h}\bar{h}^{ab} \delta\left(\xi - \xi^{\prime}\right)+ \bar{h}\bar{h}^{ab} \delta\left(\xi^{\prime} - \xi^{\prime \prime}\right)\right] \partial_{a}\left\{\delta\left(\xi - \xi^{\prime}\right)\right\}
\label{00b}
\end{eqnarray}
\begin{eqnarray}
C_{a0}^{0} = \left[\delta\left(\xi - \xi^{\prime \prime}\right)+ \delta\left(\xi^{\prime} - \xi^{\prime \prime}\right)\right] \partial_{a}\left\{\delta\left(\xi - \xi^{\prime}\right)\right\}
\label{a00}
\end{eqnarray}
\begin{eqnarray}
C_{ab}^{c} = \left[\delta\left(\xi - \xi^{\prime \prime}\right)\partial_{a}\left\{\delta\left(\xi - \xi^{\prime}\right)\right\} \delta^{c}{}_{b} + \delta\left(\xi^{\prime} - \xi^{\prime \prime}\right)\partial_{b}\left\{\delta\left(\xi - \xi^{\prime}\right)\right\} \delta^{c}{}_{a}\right] 
\label{abc}
\end{eqnarray}
and using the structure functions (\ref{00b}, \ref{a00}, \ref{abc}) we calculate the required gauge variations by applying equation (\ref{gaugevariation})
\begin{eqnarray}
\delta \lambda^{a} &=& - \dot \alpha_{a}
+ \left(\alpha_{b}\partial_{b}\lambda^{a}
 - \lambda^{b}\partial_{b}\alpha_{a} \right) + 4 \bar{h}\bar{h}^{ab}
\left(\alpha_{0}\partial_{b}\lambda_{0} - \lambda_{0}\partial_{b}
\alpha_{0}\right)\nonumber\\
\delta \lambda_{0} &=& -\dot \alpha_{0}
+\left(\alpha_{0}\partial_{a}\lambda^{a} - \lambda^{a} \partial_{a}\alpha_{0}
\right)
+\left(\alpha_{a}\partial_{a}\lambda_{0} - \lambda_{0} \partial_{a}\alpha_{a}
\right)
\label{deltalambda}
\end{eqnarray}


Now we will systematically investigate and explicitely establish the parallel between gauge symmetry and reparametrization symmetry of the p-brane actions. To start with we require to find a correspondence between the transformation parameters in both the cases. A particular usefulness of the interpolating Lagrangean formalism can be appreciated now. It is not required to find the gauge variations of $g_{ij}$ from scratch. The identification (\ref{idm}) allows us to find the required gauge variations from the corresponding transformations of $\lambda_{0}$ and $\lambda^{a}$ (\ref{deltalambda}). The Polyakov action offers the most appropriate platform to test this proposition. Indeed, the complete equivalence between the two concepts (i.e. gauge variation and reparametrisation)can be demonstrated from the Polyakov action by devising an exact mapping between the reparametrization parameters and gauge parameters by comparing the changes of $\lambda_{0}$ and $\lambda^{a}$ from the alternative approaches using the identification (\ref{idm}).
 
Under infinitesimal reparametrisation of the world volume coordinates 
\begin{eqnarray}
\xi^{\prime}{}_{i} =\xi_{i} - \Lambda_{i} 
\label{reparametrisation}
\end{eqnarray}    
where $\Lambda_{i}$ are arbitrary functions of $\xi_{i}$, the variations of the fields $X_{\mu}$ and $g_{ij}$ are
\begin{equation}
\delta X^{\mu} = \Lambda^{i} \partial_{i} X^{\mu}
= \Lambda^{0} \dot X^{\mu} + \Lambda^{a}\partial_{a}X^{\mu}
\label{deltaX}
\end{equation}
\begin{equation}
\delta g_{ij} = D_{i}\Lambda_{j} + D_{j}\Lambda_{i}
\label{deltag}
\end{equation}
where the covariant derivative is defined as usual,
\begin{equation}
D_{i}\Lambda_{j} = \partial_{i} \Lambda_{j} - \Gamma_{ij}{}^{k} \Lambda_{k}
\label{3123}
\end{equation}
with $\Gamma_{ij}{}^{k}$ being the Christoffel symbols \cite{jvh}. 

      The infinitesimal parameters $\Lambda^{i}$ characterizing reparametrization correspond to infinitesimal gauge transformations and will ultimately be related with the gauge parameters $\alpha_{k}$ introduced earlier such that the symmetry transformations on $X^{\mu}$ agree from both the approaches. Since the metric $g_{ij}$ is associated with $\lambda_{0}$ and $\lambda^{a}$ by the correspondence (\ref{idm}), equation (\ref{deltag}) will enable us to establish the complete equivalence between the variations due to the gauge transformation and the reparametrizations.
    
     To compare variations of $X^{\mu}$ from these alternative approaches we proceed as follows. From the Lagrangean corresponding to (\ref{polyakovaction}) we find
\begin{equation}
\Pi^{\mu} = - \sqrt{- g } g^{00} \dot X^{\mu} - \sqrt{- g } g^{0a}
 \partial_{a}X^{\mu}
\label{3118}
\end{equation}
Substituting $\dot X^{\mu}$ from (\ref{3118}) in (\ref{deltaX}) we get after
 some calculation
\begin{equation}
\delta X^{\mu} = \Lambda^{0} \frac{\sqrt{- g } }{g g^{00} }
 \Pi^{\mu}  + \left(\Lambda^{a} - \frac{g^{0a} }{g^{00} }
 \Lambda^{0}\right) \partial_{a}X^{\mu}
\label{3119}
\end{equation}
Now the variation of $X^{\mu}$ in (\ref{interpolatingLagrangean}) under (\ref{gaugegenerator}) is
\begin{equation}
\delta X_{\mu} = \left\{X_{\mu}, G \right\}
 = \left( \alpha_{a} \partial_{a} X_{\mu} + 2 \alpha_{0} \Pi_{\mu} \right)
\label{GM7}
\end{equation}
Comparing the above expression of $\delta X^{\mu}$ with that of (\ref{3119})
we find the mapping
\begin{eqnarray}
\Lambda^{0} =- \frac{\alpha_{0}}{\lambda_{0}}
\nonumber \\
\Lambda^{a} = \alpha_{a} -\frac{\lambda^{a}\alpha_{0}}{\lambda_{0}}
\label{3120}
\end{eqnarray}
With this mapping the gauge transformation on $X_{\mu}$ in both the formalism
agree.
The complete equivalence between the transformations can now be established by computing $\delta \lambda^{a}$ and $\delta \lambda_{0}$ from the alternative approach. The mapping (\ref{idm}) yields,
\begin{equation}
\lambda^{a} = \frac{g^{0a}}{g^{00}}
\label{GM11}
\end{equation}
We require to express these in terms of $g_{ij}$. To this end we start from the identity
\begin{equation}
g^{ij} g_{jk} = \delta^{i}{}_{k}
\label{GM12}
\end{equation}
and obtain the following equations for $\lambda^{a}$
\begin{eqnarray}
\lambda^{a} g_{ab} = - g_{0b}
\label{GM13}
\end{eqnarray}
which gives 
\begin{eqnarray}
\lambda^{a} = - \bar{g}^{ab} g_{0b}
\label{GM14}
\end{eqnarray}
Taking variation on both sides of (\ref{GM13})we get 
\begin{equation}
\delta \lambda^{a} g_{ab} = -\delta g_{0b} - \lambda^{a} \delta g_{ab} 
\label{1stvariation}
\end{equation}
Rearranging the terms conveniently we write
\begin{equation}
\delta \lambda^{a} = -\delta g_{0b} \bar{g}^{ab} - \lambda^{c} \delta g_{cb} \bar{g}^{ba}
\label{mastervariation}
\end{equation}
Now using (\ref{deltag}) we compute $\delta g_{0b}$ and $\delta g_{ab}$ and simplify using relation (\ref{GM14}) and the fact that $\bar{g}^{ab}$ is the inverse of the spatial metric $g_{ab}$. This gives the variations of $\lambda^{a}$ under reparametrization as
\begin{eqnarray}
\delta \lambda^{a} &=& - \partial_{0} \Lambda^{a} +
\lambda^{a} \partial_{0} \Lambda^{0} - \lambda^{b} \partial_{b} \Lambda^{a} 
 + \lambda^{a}\lambda^{b} \partial_{b} \Lambda^{0}
+ \Lambda^{k}\partial_{k} \lambda^{a}\nonumber \\
&& \qquad\qquad\qquad + \left(g_{00} - g_{c0}\bar{g}^{cb}g_{0b}\right)\bar{g}^{ad}\partial_{d} \Lambda^{0}
\label{GM14}
\end{eqnarray}
We further simplify the last term using a relation\footnote{This relation is obtained by substituting for $\lambda_{0}$ and $\lambda^{a}$ from (\ref{LambdaRhoeqnmotion}) in the interpolating Lagrangean (\ref{interpolatingLagrangean})}
\begin{equation}
\frac{1}{\lambda^{2}}\left[h_{00} - h_{0c}\bar{h}^{cb}h_{0b}\right] + \bar{h} = 0
\label{therelation}
\end{equation}
and (\ref{polyeqnmotion}) to get
\begin{eqnarray}
\delta \lambda^{a} &=& - \partial_{0} \Lambda^{a} +
\lambda^{a} \partial_{0} \Lambda^{0} - \lambda^{b} \partial_{b} \Lambda^{a} 
 + \lambda^{a}\lambda^{b} \partial_{b} \Lambda^{0}
+ \Lambda^{k}\partial_{k} \lambda^{a}\nonumber \\
&& \qquad\qquad\qquad + \lambda_{0}^{2}\bar{h}\bar{h}^{ad}\partial_{d} \Lambda^{0}
\label{GM15}
\end{eqnarray}
Now introducing the mapping (\ref{3120}) in (\ref{GM15}) 
we find that the variations of $\lambda^{a}$ are identical with their gauge variations in (\ref{deltalambda}). Finally we compute the variation of $\lambda_{0}$. This can be conveniently done by taking the expression of $\lambda_{0}$ in (\ref{theg}) and using the variations (\ref{deltag}). We get the expression of $\delta \lambda_{0}$ in terms of the reparametrization parametres $\Lambda_{i}$ as 
\begin{equation}
\delta \lambda_{0} = \lambda_{0} \partial_{i}\Lambda^{i} + \Lambda^{k} \partial_{k} \lambda_{0} - 2\lambda_{0} \partial_{a}\Lambda^{a} + 2 \lambda_{0} \lambda^{a} \partial_{a} \Lambda^{0}
\label{GM16}
\end{equation}
Again using the mapping (\ref{3120}) we substitute $\Lambda_{i}$ by $\alpha_{i}$ and  the resulting expression for $\delta \lambda_{0}$ agrees with that given in (\ref{deltalambda}). The complete matching, thus obtained, illustrates the equivalence of reparametrization symmetry with gauge symmetry for the generic p-brane.
\section{Conclusion}
We have discussed in the context of the bosonic p-brane how an independent metric can be generically introduced in the world volume of the brane. This has been done with the help of an interpolating action based on the first-class constraints. The specific method adopted here leads to the introduction of the metric in a very special way, namely we have achieved a segregation of the $\left(p + 1\right)$ dimentional world volume in the $p$ dimentional spatial part and the Lagrange multipliers analogous to the lapse and shift variables of classical gravity. Using this correspondence we have shown that the Arnowitt--Deser--Misner like decomposition of the brane world volume emerges from our analysis. A comprehensive analysis of the gauge symmetries of the interpolating action has been elaborated and equivalence of gauge and reparametrisation invariances has been established. 

\section*{Acknowledgment}
RB acknowledges support from the 
Alexander von Humboldt Foundation. He also thanks the membres of the 
Max-Planck-Institut f$\ddot{u}$r Physik, F$\ddot{o}$hringer Ring 6,80805 M$\ddot{u}$nchen, Germany and Universit$\ddot{a}$t M$\ddot{u}$nchen, Fauklt$\ddot{a}$t f$\ddot{u}$r Physik, Theresienstr. 37, 8033 M$\ddot{u}$nchen, Germany for their hospitality, where part of the work is done.
PM likes to thank the Director, S. N. Bose National Centre for Basic Sciences for the award of visiting associateship. AS wants to thank the Council of Scientific and Industrial Research (CSIR), Govt. of India, for financial support and the Director, S. N. Bose National Centre for Basic Sciences, for providing computer facilities. 



\end{document}